\documentclass{article}
\usepackage{epsfig}
\begin{document}
\begin{center}
{\LARGE \bf Berry's phase in noncommutative  spaces }
\end{center}
\vspace{2.cm}
\begin{center}
{\LARGE\bf S. A. Alavi}
\end{center}
\begin{center}
\textit{High Energy Physics Division, Department of Physics,
University of
Helsinki\\
and\\
Helsinki Institute of Physics, FIN-00014 Helsinki, Finland.\\
On leave of absence from : Department of Physics, Ferdowsi
University of
Mashhad, Mashhad, P. O. Box 1436, Iran\\
E-mail: $Ali.Alavi@Helsinki.fi$\\
$s_{-}alialavi@hotmail.com$}
\end{center}

\textbf{1 Abstract.} \textsl{We introduce the perturbative aspects
of noncommutative quantum mechanics. Then we study the Berry's
phase in the framework of noncommutative quantum mechanics.
The results show deviations from the usual quantum mechanics which depend on
the parameter of space/space noncommtativity.}\\

\textbf{1 Introduction}.\\

Noncommutative quantum mechanics have received a wide attentian
once it was realized that they could be obtained as low energy
limit of string theory in the presence of a $B$ field [1,2]. In
field theories the noncommutativity is introduced by replacing the
standard product by the star product. for a manifold parameterized
by the coordinates $x_{i}$, the noncommutative relation can be
written as :
\begin{equation}
[\hat{x}_{i},\hat{x}_{j}]=i\theta_{ij}\hspace{1.cm}[\hat{x}_{i},\hat{p}_{i}]=i\delta
_{ij}\hspace{1.cm} [\hat{p}_{i},\hat{p}_{j}]=0 .
\end{equation}
Many physical problems have been studied in the framework of the
noncommutative quantum mechanics (NCQM), see e.g. [6-19]. NCQM is
formulated in the same way as the standard quantum mechanics SQM
(quantum mechanics in commutative spaces), that is in terms of the
same dynamical variables represented by operators in a Hilbert
space and a state vector that evolves according to the
Schroedinger equation :
\begin{equation}
i\frac{d}{dt}|\psi>=H_{nc}|\psi> ,
\end{equation}
we have taken in to account $\hbar=1$. $H_{nc}\equiv H_{\theta}$
denotes the Hamiltonian for a given system in the noncommutative
space. In the literatures two
approaches have been considered  for constructing the NCQM :\\
a) $H_{\theta}=H$, so that the only difference between SQM and
NCQM is the presence of a nonzero $\theta$ in the commutator of
the position operators i.e. equ.(1). \\
b) By deriving the Hamiltonian from the moyal analog of the
standard Schroedinger equation :
\begin{equation}
i\frac{\partial}{\partial t}\psi(x,t)=H(p=\frac{1}{i}\nabla,x)\ast
\psi (x,t)\equiv H_{\theta}\psi(x,t) ,
\end{equation}
where $H(p,x)$ is the same Hamiltonian as in the standard theory,
and as we observe the $\theta$ - dependence enters now through the
star product [17]. In [19], it has been shown that these two
approaches lead to the same physical theory. For the Hamiltonian
of the type :
\begin{equation}
H(\hat{p},\hat{x})=\frac{\hat{p}^{2}}{2m}+V(\hat{x}) .
\end{equation}
The modified Hamiltonian $H_{\theta}$ can be obtained by a shift
in the argument of the potential [6,17] :
\begin{equation}
x_{i}=\hat{x}_{i}+\frac{1}{2}\theta_{ij}\hat{p}_{j}\hspace{2.cm}\hat{p}_{i}=p_{i}
.
\end{equation}
which lead to
\begin{equation}
H_{\theta}=\frac{p^{2}}{2m}+V(x_{i}-\frac{1}{2}\theta_{ij}p_{j}) .
\end{equation}
The variables $x_{i}$ and $p_{i}$ now, satisfy in the same
commutation relations as the usual case :
\begin{equation}
[x_{i},x_{j}]=[p_{i},p_{j}]=0\hspace{1.cm}[x_{i},p_{j}]=\delta_{ij} .
\end{equation}

Topological effects are among the most important quantum phenomena
discovered since the creation of the quantum mechanics. The
Aharonov-Bohm (AB) effect [3], was the first such effect which
changed basic assumptions about the role of the potentials in
physics. There is also a geometric effect discovered by Berry [4],
almost two decades later. Berry investigated quantum systems that
undergo an adiabatic evolution goverened by several changing
parameters. When these parameters return to their initial values,
quantum systems return to their initial states up to a phase.
Berry found that the phase contains a geometric part, which
depends on the path in the space of the parameters, but not on the
rate evolution of the system along the path. Aharonov and Anandan
subsequently generalized Berry's phase to nonadiabatic evolution
[5]. The AB effect in noncommutative spaces has been studied in
[7] and [8]. In this paper we study the Berry's phase in
noncommutative
spaces.\\

\textbf{2 Perturbation aspects of noncommutative dynamics. }\\

In this section we discuss the "perturbation aspects of
noncommutative dynamics". The perturbation aspects of $q$-deformed
dynamics in one dimensional $q$-spaces has been studied in [20]
and [21]. Using
\begin{equation}
U(x+\Delta x)=U(x)+\sum_{n=1}^{\infty}  \frac{U^{(n)}(x)}{n!} (\Delta x)^{n}
,
\end{equation}
and equ.(6) for small $\theta$ we have :
\begin{equation}
H_{nc}=\frac{p^{2}}{2m}+V(x_{i})+\sum_{n=1}^{\infty}
\frac{V^{(n)}(x_{i})}{n!} (\Delta x_{i})^{n} ,
\end{equation}
where $\Delta x_{i}=-\frac{1}{2}\theta \epsilon_{ij}p_{j} $ and
$H=\frac{p^{2}}{2m}+V(x)$ is the Hamiltonian in ordinary(commutative) space.
To the first order we have :
\begin{equation}
H_{nc}=\frac{p^{2}}{2m}+V(x_{i})+\Delta x_{i} \frac{\partial
V}{\partial x_{i}}=H+\Delta x_{i}\frac{\partial
V}{\partial x_{i}}=H+\theta H_{I} .
\end{equation}

We can use perturbation theory to
obtain the eigenvalues and eigenfunctions of $H_{nc}$ :
\begin{equation}
E_{n}=E^{0}_{n}+\Delta E^{0}_{n}=E^{0}_{n}+\theta E^{(1)}_{n}+\theta^{2}
E^{(2)}_{n}+...\hspace{.3cm} .
\end{equation}
\begin{equation}
\psi_{n}=\phi_{n}+ \sum_{k\neq n}C_{nk}(\theta)\phi_{k} .
\end{equation}
where :
\begin{equation}
C_{nk}(\theta)=\theta C^{(1)}_{nk}+\theta^{2} C^{(2)}_{nk}+...\hspace{.3cm}
.
\end{equation}
To the first order in perturbation theory we have :
\begin{equation}
\theta E^{(1)}_{n}= <\phi_{n}|\theta H_{I} |\phi_{n}> ,
\end{equation}
\begin{equation}
\psi_{n}=\phi_{n}+\theta \sum_{k\neq n}C^{(1)}_{nk}\phi_{k} ,
\end{equation}
\begin{equation}
\theta C^{(1)}_{nk}=\frac{<\phi_{k}|\theta
H_{I}|\phi_{n}>}{E^{0}_{n}-E_{k}^{0}} ,
\end{equation}
where $E^{0}_{n}$ and $\phi_{n}$ are the $n$th eigenvalue and
eigenfunction of the Hamiltonian $H$. $E_{n}$ and $\psi_{n}$ are
the $n$th eigenvalue and eigenfunction of $H_{nc}$. For example
for the Hydrogen atom :
\begin{equation}
V(r)=-\frac{ze^{2}}{r}=-\frac{ze^{2}}{\sqrt{\hat{x}\hat{x}}} .
\end{equation}
and we have :
\begin{equation}
\theta H_{I}=-\frac{1}{2}\theta
\epsilon_{ij}p_{j}(\frac{x^{i}ze^{2}}{r^{3}})=-ze^{2}\theta\frac{x_{i}\epsilon_{ij}p_{j}}{2r^{3}}
.
\end{equation}
which is in agreement with the result of Ref. [6]. Using (15) and
(16) we can find the eigenfunctions for the Hydrogen atom in
noncommutative spaces.
\begin{equation}
\psi_{n}^{Hydrogen}=\phi_{n}+ \sum_{k\neq n}\frac{<\phi_{k}|
\theta H_{I}|\phi_{n}>}{E^{0}_{n}-E^{0}_{k}}\phi_{k} ,
\end{equation}
where $\theta H_{I}$ is given by equ.(18).\\

\textbf{2 Berry's phase in noncommutative spaces. }\\

Now the Hamiltonian $H_{nc}$, its eigenfunctions $\psi_{n}$ and
eigenvalues $E_{n}$ are known (equs. (11)-(13)) and we can
calculate the Berry's phase in noncommutative case. suppose a
quantum system is in $n$th eigenstate :
\begin{equation}
H_{nc}|\psi_{n}>=E_{n}\psi_{n} .
\end{equation}
If $H_{nc}$ is slowly changed, then according to the adiabatic
theorem the system remains in the $n$th eigenstate of the
slowly-changing Hamiltonian $H_{nc}$. If at time $\tau$ it returns
to its initial form $H_{nc}(\tau)=H_{nc}(t=0)$, it follows that
the system must return to its initial state up to a phase factor,
and we have :
\begin{equation}
|\psi_{n}(\tau)>=e^{-i\int_{0}^{\tau}E_{n}(t^{\prime})d
t^{\prime}}e^{i\eta_{n}(\tau)}|\psi_{n}> .
\end{equation}
>From the Schroedinger equation we have :
\begin{equation}
i\frac{\partial |\psi_{n}>}{\partial t}=E_{n}(t)|\psi_{n}(t)>.
\end{equation}
Differentiating eq.(21) at time $t$, and plugging in to the
Schroedinger equation, we obtain :
\begin{equation}
\frac{\partial}{\partial t}|\psi_{n}(t)>+i\frac{d\eta_{n}(t)}{d
t}|\psi_{n}(t)>=0.
\end{equation}
Multiplying on the left by $<\psi_{n}(t)|$, we have :
\begin{equation}
\eta_{n}(\tau)=i\int_{0}^{\tau}<\psi_{n}(t)|\frac{\partial}{\partial
t }|\psi_{n}(t)> d t  .
\end{equation}
If the time dependence of the Hamiltonian arises because we are
changing some prameters $R$ with time, then we may write
\begin{equation}
\eta_{n}(\tau)=i\int_{0}^{\tau}<\psi_{n}(t)|\frac{\partial}{\partial
t }|\psi_{n}(t)> d t=i\oint dR
<\psi_{n}(R)|\nabla_{R}\psi_{n}(R)> .
\end{equation}
By substituting for $\psi_{n}(t)$ for a given time $t$ :
\begin{equation}
|\psi_{n}(t)>=|\phi_{n}(t)>+\sum_{k\neq n} C_{nk}(\theta)
|\phi_{k}(t)> .
\end{equation}

we have : \\

$<\psi_{n}(t)|\frac{\partial}{\partial
t}|\psi_{n}(t)>=$

\begin{equation}
[<\phi_{n}(t)|+\sum_{k\neq n} C_{nk}(\theta)
<\phi_{k}(t)|]\frac{\partial}{\partial t}[|\phi_{n}(t)>+ \sum_{\ell\neq n}
C_{n\ell}(\theta)
|\phi_{\ell}(t)>] .
\end{equation}

which leads to :\\

$\eta_{n}(\tau)=i\int_{0}^{\tau} dt<\phi_{n}(t)|\frac{\partial}{\partial
t}|\phi_{n}(t)>+$\\

$i\sum_{i=1}^{\infty}\theta^{i}\int_{0}^{\tau} dt\{ \sum_{\ell\neq
n}C^{(i)}_{n\ell}<\phi_{n}(t)|\frac{\partial}{\partial
t}|\phi_{\ell}(t)>+\sum_{k\neq
n}C^{(i)}_{nk}<\phi_{k}(t)|\frac{\partial}{\partial
t}|\phi_{n}(t)>$

\begin{equation}
+ \sum_{r=1}^{\infty}\sum_{\ell\neq
n}\sum_{k\neq
n}\theta^{r}
C_{n\ell}^{(i)}C_{nk}^{(r)}<\phi_{\ell}(t)|\frac{\partial}{\partial
t}|\phi_{k}(t)>\} .
\end{equation}
or :\\

$\eta_{n}=i\oint dR<\phi_{n}(R)|
\nabla_{R}\phi_{n}(R)>+$\\

$i\sum_{i=1}^{\infty}\theta^{i}\oint dR\{ \sum_{\ell\neq
n}C^{(i)}_{n\ell}<\phi_{n}(R)|
\nabla_{R}\phi_{\ell}(R)>+\sum_{k\neq
n}C^{(i)}_{nk}<\phi_{k}(R)|
\nabla_{R}\phi_{n}(R)>$

\begin{equation}
+ \sum_{r=1}^{\infty}\sum_{\ell\neq
n}\sum_{k\neq
n}\theta^{r} C_{n\ell}^{(i)}C_{nk}^{(r)}<\phi_{\ell}(R)|
\nabla_{R}\phi_{k}(R)>\} .
\end{equation}

The first term is Berry's phase in commutative space and the
second term gives the correction to the Berry's phase due to the
noncommtativity of space. If the parameter $\theta$ is small, the higher
order
terms (higher powers in $\theta$) are very small and we have : \\

$\eta_{n}=i\oint dR<\phi_{n}(R)|
\nabla_{R}\phi_{n}(R)>+ $

\begin{equation}
i\theta\oint dR\{\sum_{\ell\neq
n}C^{(1)}_{n\ell}<\phi_{n}(R)|
\nabla_{R}\phi_{\ell}(R)>+\sum_{k\neq
n}C^{(1)}_{nk}<\phi_{k}(R)|
\nabla_{R}\phi_{n}(R)>\}+O(\theta^{2}) .
\end{equation}
As we mentioned above the argument can be generalized to nonadiabatic
evolution.\\
Let us consider the case of a spin-$\frac{1}{2}$ particle in a
magnetic field. If the magnetic field $\mathbf{B}(t)$ initially
lie along the $z$-direction $ \mathbf{B}(t=0)=\mathbf{B}\hat{z}$,
then the initial Hamiltonian is given by :
\begin{equation}
H(t=0)=-\mathbf{M}\mathbf{.}
\mathbf{B}=\frac{eg}{4mc}\mathbf{\sigma}\mathbf{.}
\mathbf{B}=\frac{eg}{2mc}B S_{z} .
\end{equation}
Assume that the particle initially starts out in an eigenstate of
$S_{z}$, we want to  determine what happens to the particle if the
direction of the $B$-field is rotated in some matter such that at
time $t=\tau$, the field again lines up with the $z$- axis. If the
$B$-field doesn't depend on the coordinates i.e. it is constant in
space, then $H_{I}=0$ and we have :
\begin{equation}
H_{nc}=H .
\end{equation}
This means that for a spin-$\frac{1}{2}$ particle at the presence
of constant magnetic field in a noncommutative space, the
correction to the Berry's phase vanishes.\\
\textbf{Acknowledgment.}\\
I would like to thank Professor Masud Chaichian for his warm
hospitality during my visit to the university of Helsinki. I am very
grateful to M. M. Sheikh- Jabbari for his useful comments. I acknowledge
S. M. Harun-or-Rashid for his help on preparation of the manuscript.
This work is partialy supported by
the Ministry of Scince, Research and Technology of Iran.\\
\textbf{References.}\\
1. N. Seiberg and E. Witten, JHEP9909, 032(1999). \\
2. D. Bigahil, L. Susskind, Phys. Rev. D, 62 (2000)066004. \\
3. Y. Aharonov and D. Bohm, Phys. Rev. 11, (1959) 485.\\
4. M. V. Berry, Proc. Roy. Soc. A392, 45 (1984).\\
5. Y. Aharonov and J. Anandan, Phys. Rev. Lett. 58, 1593 (1987).\\
6. M. Chaichian, M. M. Sheikh-Jabbari and A. Tureanu, Phys. Rev.
Lett. 86, 2716 (2001).\\
7. M. Chaichian, A. Demichev, P. Pre$\breve{s}$najder, M. M.
Sheikh-Jabbari, and A. Tureanu, Phys. Lett. B 527 (2002) 149.\\
8. M. Chaichian, A. Demichev, P. Pre$\breve{s}$najder, M. M.
Sheikh-Jabbari, and A. Tureanu, Nucl. Phys. B611(2001) 383.\\
9. A. Smailagic and E. Spallucci, Phys. Rev. D 65, 107701
(2002).\\
10. $\ddot{O}$. Dayi and A. Jellal, Phys. Lett. A 287, 349
(2001).\\
11. P-M Ho and H-C Kao, Phys. Rev. Lett 88, 151 602 (2001).\\
12. R. Banerjee, hep-th/0106280. \\
13. J. Gamboa, F, M$\acute{e}$ndez, M. Loewe and J. C. Rojac, Mod.
Phys. Lett. A 16, 2075 (2001).\\
14. V. P. Nair and A. P. Polychronakos, Phys. Lett. B 505, 267
(2001).\\
15. B. Muthukamar and R. Mitra, hep-th/0204149.\\
16. D. Kochan and M. Demetrian, hep-th/0102050.\\
17. L. Mezincesu, hep-th/0007046.\\
18. Y. Zunger, JHEP 0104, 039 (2001).\\
19. O. Espinosa and P. Gaete, hep-th/0206066.\\
20. Jian-zu Zhang and P. Osland. Eur. Phys. J. C20(2001) 393\\
21. S. A. Alavi, hep-th/0205229

\end{document}